\documentstyle[preprint,aps]{revtex}
\tightenlines 
\begin{document}
\draft

\title{Dissipative Interaction and Anomalous Surface Absorption \\of Bulk
Phonons at a Two-Dimensional Defect in a Solid}

\author{Yu. A. Kosevich\thanks{Corresponding author. On leave from: Moscow State Technological
University ``STANKIN'', 101472 Moscow, Russia; electronic address: 
yukos@mpipks-dresden.mpg.de}} 

\address{Max Planck Institute for Physics of Complex Systems, Noethnitzer Str.
38, 
D-01187 Dresden, Germany;\\ Tel.: +49 351 871 1216; Fax: +49 351 871 2199; 
E-mail: yukos@mpipks-dresden.mpg.de}

\author{E. S. Syrkin} 
\address{B.I.Verkin Institute for Low Temperature Physics, 310164 Kharkov, Ukraine}

\maketitle
\begin{abstract}

We predict an extreme sensitivity to the dissipative losses 
of  the resonant  interaction 
of bulk phonons with a 2D defect in a solid.  
We show that the
total resonant reflection of the transverse phonon at the 2D defect, 
described earlier without an account for dissipation,
occurs only in the limit of extremely weak dissipation and is changed into
almost total transmission by relatively weak bulk 
absorption. Anomalous surface absorption of the transverse phonon, when 
one half of the incident acoustic energy is absorbed at the 2D defect,  
is predicted for the case of ``intermediate" bulk dissipation.
\end{abstract}


\newpage   

 Two-dimensional (planar) defects in the bulk of a solid, such as stacking faults, grain and twin
boundaries, dislocation walls, subtantially modify the 
vibrational 
spectrum of the system. Particularly strong effects can occur at resonant interaction of bulk  
phonons with a 2D defect in a solid, namely the total reflection of the grazing
acoustic wave at an ultrathin (2D) embedded layer [1,2], 
the total transmission at resonance with intrinsic
dynamical degrees of freedom of a 2D defect layer [3,4], 
the total reflection of a bulk phonon at resonance with an 
asymmetric vibrational mode of an $inhomogeneous$ 2D elastic layer with complex
internal structure [5], and the total 
reflection of the transverse acoustic wave at resonance with the quasilongitudinal
leaky wave  of a $homogeneous$ 
2D elastic layer [6,7]. To 
elucidate the possibilities of the experimental observation of these wave 
phenomena, caused by relatively thin elastic layer sandwiched in a solid, 
one has  to study the
effect on them of other relatively small interactions. In such a way one can
check whether these phenomena can ``survive" in a real
experimental system.

In the present work we analyse the effect of bulk and surface 
dissipation  on the interaction of a transverse elastic 
wave (acoustic phonon) with a homogeneous 2D elastic defect layer in a solid. 
The interest
in this study is stipulated by the fact that three 
characteristic lengths simultaneously appear in the problem, namely the 
wavelength, the effective thickness of the 2D defect, and the
effective ``dissipative length" (or effective mean free path of the bulk or
surface excitations which cause the dissipation). We show that the fine
interplay between these three lengths determines the resonant interaction 
of bulk phonons with the planar defect. We emphasize that a   
relatively weak bulk and surface dissipation substantially effects 
the total reflection of a transverse bulk phonon at a 2D defect 
layer at resonance with the quasilongitudinal pseudosurface (leaky)  
wave at the layer,
described in Refs. [6,7] without an account for the dissipation. 
We show that the (bulk and surface)  
dissipation causes finite transmission of bulk phonons 
through the 2D defect for any angle of incidence, including the resonant
one. This result is well consistent with the statement, 
known from the theory of the Kapitza thermal resistance, that generally the
dissipation  enhances the phonon transmission through the interface 
between two media (see, e.g., Refs. [8,9]). We predict that at resonance with
the leaky wave the 
transmission and reflection amplitudes are determined mainly 
by the dimensionless ratio
of the longitudinal bulk dissipative length $a_{l}$ and effective thickness
$a^{\star}$ of the 2D defect   
which is enhanced by the large factor $(a^{\star}k_{x})^{-2}\gg 1$, where
$k_{x}$ is the tangential component of the incident wavevector. 
This characteristic ratio 
is proportional to the bulk viscosity and thermal conductivity of the solid and 
is inversely proportional to the square of the frequency. 
Almost total reflection of
the transverse bulk phonon at the 2D defect occurs only in the limit of 
extremely  weak dissipation  when  this ratio is much less than 
unity. This limit in most cases is hard to reach experimentally. 
When this ratio has the order of 
or is larger than unity, the total resonance reflection of the transverse bulk
phonon is changed into almost total transmission 
(for all angles of incidence, including the resonant
one). The origin of such strong effect of the dissipation on resonance 
interaction of bulk phonons with a 2D defect is the 
large penetration depth of the  
quasilongitudinal leaky wave and the strong enhancement of its  
amplitude at resonance with the incident transverse phonon (cf. Refs. [6,7]). 
It causes an extremely strong enhancement of the 
surface absorption of the incident wave. 
The enhancement of surface
absorption of an acoustic phonon incident from liquid helium at 
resonance with the leaky
Rayleigh wave at the liquid helium - solid interface has been 
described theoretically [10] and investigated
experimentally (for a recent review, see Ref. [11]). In the case of a 2D defect layer sandwiched 
in a solid, the effect of bulk dissipation on resonance reflection 
is strongly enhanced in
comparison with the interface between liquid and solid 
due to the presence of an 
additional small and frequency-dependent 
parameter which governs the interaction between the
incident transverse phonon and the quasilongitudinal leaky wave, namely the
parameter $a^{\star}k_{x}\ll 1$ which is the ratio 
between the effective thickness of the layer and the wavelength.  
We show that  the coefficient of
surface absorption, which follows from the reflection and transmission
coefficients with an account for the bulk and surface dissipation, 
exactly coincides with the
coefficient of surface absorption which can be obtained directly with the use 
of the dissipative function of a solid. This confirms the consistency 
of the proposed description  of the effect of bulk and surface 
dissipation on resonant 
reflection and refraction of elastic 
waves in dissipative media, which in general is a rather tough problem (see, e.g.,
Ref. [12]).  We describe also the conditions for 
anomalous surface absorption of the transverse phonon at the 2D defect when one half of
the incident acoustic energy is absorbed at the ultrathin elastic layer.  
This effect, which can be interesting from the experimental point of 
view, is similar to anomalous surface 
absorption of the grazing shear elastic wave in a thin gap between two solids 
caused by the dissipative van der Waals interaction between the solids [13]. 
In both cases,  the anomalous
surface (interface) absorption of the bulk phonon is stipulated by resonant
dissipative 
interaction with the leaky elastic wave. 
      
At first, we consider the interaction of the bulk
transverse wave with frequency $\omega$ with a 
2D defect placed on the $z=0$ plane in an 
isotropic elastic solid without an account for the bulk and surface
dissipation. In an isotropic
solid we can introduce 
scalar and vector potentials $\varphi$ and $\vec{\psi}$, which satisfy the
equations of motion $\ddot{\varphi}=c_{l}^2\Delta\varphi$,  
$\ddot{\vec{\psi}}=c_{t}^2\Delta\vec{\psi}$ and describe the longitudinal and transverse
elastic displacements in the
solid, $\vec{u}={\bf grad}\varphi+{\bf curl}\vec{\psi}$, where $c_{l}$ and
$c_{t}$ are, respectively, longitudinal and transverse sound velocities. 
In the case when the
plane of incidence coincides with the $XOZ$ plane, the vector potential can
be taken as $\vec{\psi}=(0,\psi,0)$. 
For the transverse wave incident from the $z<0$ semispace and polarized in the
plane of incidence, the wave fields in
the medim 1 ($z<0$) and medium 2 ($z>0$) have the following form:
\begin{eqnarray}
\psi^{(1)}&=&\left(e^{ik_{t}\cos\theta_{t}z}+
r_{t}e^{-ik_{t}\cos\theta_{t}z}\right)e^{ik_{t}\sin \theta_{t}x-i\omega t},\\
\varphi^{(1)}&=&
r_{l}e^{-ik_{l}(\cos\theta_{l}z+\sin \theta_{l}x)-i\omega t},\\
\psi^{(2)}&=&
d_{t}e^{ik_{t}(\cos\theta_{t}z+\sin \theta_{t}x)-i\omega t},\\
\varphi^{(2)}&=&
d_{l}e^{ik_{l}(\cos\theta_{l}z+\sin \theta_{l}x)-i\omega t},
\end{eqnarray}
where $\theta_{t}$ is the angle of incidence of the transverse wave, 
$\theta_{l}$ is the angle of reflection and transmission of the 
longitudinal wave,  
$r_{t,l}$ and $d_{t,l}$ are the reflection and
transmission amplitudes for the transverse and longitudinal waves,  
respectively, $k_{t,l}=\omega/c_{t,l}$ are bulk wavenumbers.  

For the macroscopic description of the interaction of an elastic wave with 
a planar  
defect with the effective thickness much less than the wavelength, one can
make use of a set of effective boundary (matching) conditions for the elastic
displacements and stresses at the plane of the 2D defect (see, e.g.,
Refs. [1,2,5,14]). In the case of a homogeneous 2D defect which does not possess complex
internal structure and intrinsic dynamical degrees of freedom, these boundary
conditions can be reduced in the simplest case 
to the continuity of the elastic displacements $u_{i}$ and to the discontinuity of
surface-projected bulk stresses $\sigma_{zi}$, $i=1,2,3$ (for more general
boundary conditions which account for the discontinuity both of
surface-projected bulk stresses and elastic displacements, see
Refs. [5,15]). For the problem under consideration, the following
boundary conditions will be used:
\begin{eqnarray}
u_{x}^{(1)}-u_{x}^{(2)}&=&0, ~~~ u_{z}^{(1)}-u_{z}^{(2)}=0,\\ 
\sigma_{zz}^{(1)}-\sigma_{zz}^{(2)}&=&
g_{1}\displaystyle\frac{\partial^{2}u_{z}}{\partial x^{2}}
-\varrho_{s}\displaystyle\frac{\partial^{2}u_{z}}{\partial t^{2}}, \\
\sigma_{zx}^{(1)}-\sigma_{zx}^{(2)}&=&\tilde
h_{11} \displaystyle\frac{\partial^{2}u_{x}}{\partial x^{2}}
-\varrho_{s}\displaystyle\frac{\partial^{2}u_{x}}{\partial t^{2}},
\end{eqnarray}
where $\varrho_{s}$ is the total (positive) mass of the defect layer per unit 
surface area, $g_{1}=g_{xx}$ and $h_{11}=h_{xxxx}$ are, 
respectively, the components of the tensors of the 
surface (interface) stresses and elastic moduli, $\tilde h_{11}=h_{11}+g_{1}$. 

For the components of the elastic displacements and stresses in an isotropic
solid which enter the 
boundary conditions (5)-(7), one has the following expressions in terms of
scalar and vector potentials: 
\begin{eqnarray}
u_{x}&=&\displaystyle\frac{\partial \varphi}{\partial x}-\displaystyle
\frac{\partial \psi}{\partial z},  
\quad u_{z}=\displaystyle\frac{\partial \varphi}{\partial z}+\displaystyle
\frac{\partial \psi}{\partial x},\\ 
\sigma_{zz}&=&2\rho c_{t}^{2}\left(\displaystyle\frac{\partial^{2} \varphi}
{\partial z^{2}}+\displaystyle
\frac{\partial^{2} \psi}{\partial x \partial z}\right)+
\rho (c_{l}^{2}-2c_{t}^{2})
\left(\displaystyle\frac{\partial^{2} \varphi}
{\partial x^{2}}+
\displaystyle\frac{\partial^{2} \psi}
{\partial z^{2}}\right), \\
\sigma_{zx}&=&\rho c_{t}^{2}\left(2\displaystyle\frac{\partial^{2} \varphi}
{\partial x \partial z}+\displaystyle
\frac{\partial^{2} \psi}
{\partial x^{2}}-
\displaystyle\frac{\partial^{2} \psi}
{\partial z^{2}}
\right), 
\end{eqnarray}
where $\rho$ is density of the solid. 

Making use of Eqs. (1)-(10), we find  the following expressions for the 
reflection and transmission amplitudes:
\begin{eqnarray}
r_{t} &=&[2i\rho k_{t}\cos\theta_{l}
\left(\varrho_{sz}\sin^{2}\theta_{t}-
\varrho_{sx}\cos^{2}\theta_{t}\right)-
\varrho_{sz}\varrho_{sx}k_{l}k_{t}\cos(\theta_{l}
-\theta_{t})\cos(\theta_{l}+\theta_{t})]
\displaystyle\frac{1}{\Delta}, \\
r_{l} &=&-[k_{l}k_{t}\varrho_{sx}\varrho_{sz}\cos(\theta_{l}-
\theta_{t})+i\rho(\varrho_{sz}k_{l}\cos\theta_{l}+
\varrho_{sx}k_{t}\cos\theta_{t})]
\displaystyle\frac{2\sin\theta_{l}\cos\theta_{t}}{\Delta}, \\
d_{t}&=&[2\rho\cos\theta_{l}-i\varrho_{sx}k_{l}
\sin^{2}\theta_{l}-i\varrho_{sz}k_{l}\cos^{2}\theta_{l}]
\displaystyle\frac{2\rho\cos\theta_{t}}{\Delta}, \\
d_{l}&=&[
\varrho_{sz}k_{l}\cos\theta_{l}-\varrho_{sx}k_{t}
\cos\theta_{t}]
\displaystyle\frac{2i\rho\sin\theta_{l}
\cos\theta_{t}}{\Delta}, \\
\Delta&=&\left[2\rho\cos\theta_{t}-
i\varrho_{sz}k_{l}\cos(\theta_{l}-\theta_{t})\right]\left[2\rho\cos\theta_{l}-
ik_{t}\varrho_{sx}\cos(\theta_{l}-\theta_{t})
\right],
\end{eqnarray} 
where the conservation law of the tangential component of the 
phonon momentum has been used , 
\begin{equation}
k_{x}=k_{t}\sin\theta_{t}=k_{l}\sin\theta_{l}, 
\end{equation}
and the following notations are introduced:
\begin{eqnarray}
\varrho_{sx}&=&\varrho_{s}
-\tilde{h}_{11}\displaystyle\frac{k_{x}^{2}}{\omega^{2}}, \\
\varrho_{sz}&=&\varrho_{s}-g_{1}
\displaystyle\frac{k_{x}^{2}}{\omega^{2}}.
\end{eqnarray}

From Eqs. (11)-(15) follows that in general one has $r_{t}\sim r_{l}\sim 
d_{l}\sim k_{t}a^{\star}\ll 1$, $d_{t}\sim 1$, which corresponds to   
almost total transmission of the wave through the thin (2D) elastic layer with 
effective thickness $a^{\star}\sim\varrho_{sx}/\rho$ much less that the
wavelength $\lambda$ [16]. But for the resonant (non-grazing)  
angle of incidence $\theta_{t}=\theta_{t}^{(r)}$, 
which corresponds to $\cos\theta_{l}^{(r)}\sim ik_{x}a^{\star}$,  
the reflection amplitude substantially increases and  reaches the value of
unity, $r_{t}=1$, while the transmission coefficient simultaneously turns into
zero (cf. Refs. [6,7]).   Indeed, for the
``overcritical" angles of incidence $\theta_{t}>\arcsin
(c_{t}/c_{l})$, when the $\cos\theta_{l}$ is imaginary,  
one can introduce the inverse decay length $\kappa$ for the
exponential decay of the 
longitudinal wave field away from the planar defect: 
$\cos\theta_{l}=i\kappa c_{l}/\omega$. Then from the 
requirement of the total reflection $d_{t}=0$, 
from Eq. (13) we find the following equation for the
parameter $\kappa$:
\begin{equation}
2\rho\kappa-
\varrho_{sx}k_{x}^{2}+\varrho_{sz}\kappa^{2}=0. 
\end{equation}

This equation in general has two roots for the real parameter $\kappa$, which 
have the following form in the longwavelength limit $k_{x}\ll\rho/\sqrt{\mid
\varrho_{sx}\varrho_{sz}\mid}$:
\begin{eqnarray}
\kappa_{1}&=&\displaystyle\frac{1}{2}\frac{\varrho_{sx}}{\rho} k_{x}^{2}=
\displaystyle\frac{\varrho_{s}}{2\rho}\left[1-\frac{c_{l}^{(s) 2}}{c_{l}^{2}}\right]
k_{x}^{2}, \\
\kappa_{2}&=&-2\displaystyle\frac{\rho}{\varrho_{sz}}=
-\displaystyle\frac{2\rho\omega^{2}}{\varrho_{s}\omega^{2}-
g_{1}k_{x}^{2}}.
\end{eqnarray}

In the case of $\varrho_{sx}>0$, the first root $\kappa_{1}\ll k_{x}$ (20) corresponds to the
pseudosurface (leaky) wave with $\omega^{2}\approx c_{l}^{2}k_{x}^{2}$ and 
almost longitudinal polarization, 
which can propagate along the planar defect with locally decreased longitudinal
velocity $c_{l}^{(s)}=\sqrt{\tilde h_{11}/\varrho_{s}}<c_{l}$ 
(see Refs. [17,14,1,2,6,7]). 
If one neglects dissipation, resonant excitation of this 
quasilongitudinal leaky wave 
causes total reflection of the bulk transverse wave by the planar defect [6,7]. 
The second
root $\kappa=\kappa_{2}$  (21) is unphysical one for homogeneous planar defect with 
finite surface mass  $\varrho_{s}$. It is worth  mentioning that 
Eq. (19) can be
also obtained from the solution of the problem of the total ``nontransmission"
of
the transverse acoustic wave through the planar defect for the overcritical
angle of incidence $\theta_{t}$, when the potentials 
have the following form:
\begin{eqnarray} 
\psi^{(1)}&=&( e^{iqz}+r_{t}e^{-iqz})
e^{ik_{x}x-i\omega t}, ~~
\varphi^{(1)}=r_{l}e^{\kappa z+ik_{x}x-i\omega t},\\
\psi^{(2)}&=&0, ~~
\varphi^{(2)}=d_{l}e^{-\kappa z+ik_{x}x-i\omega t},
\end{eqnarray}
where 
$q=\sqrt{\omega^{2}/c_{t}^{2}-k_{x}^{2}}=k_{t}\cos\theta_{t}$. 
One can show that the wave fields (22) and (23) satisfy boundary conditions 
(5)-(7), together with the relations (8)-(10),
 if the parameter $\kappa$ is a solution of Eq. (19) and the
reflection amplitude $r_{t}$ in Eq. (22) has the unit modulus, 
$r_{t}=\exp(i\phi_{r})$,  
which reflects the conservation law of the acoustic energy flux 
through the planar defect.  With the use of the above notations 
for $\kappa$ and $q$, from Eq. (11) we find the following expression 
for the reflection amplitude $r_{t}$ under the condition (19):
\begin{equation}
r_{t}=\frac{\varrho_{sx}q+i\varrho_{sz}\kappa}
{\varrho_{sx}q-i\varrho_{sz}\kappa}.
\end{equation}
For the value of $\kappa=\kappa_{1}$ given by Eq. (20), from Eq. (24) 
follows  that $r_{t}\approx 1$ (or $\phi_{r}\ll 1$) since $\kappa\ll (k_{x}, q)$. 

In a similar way we can also investigate the total transmission of the transverse 
acoustic wave through the 2D defect. From the requirement $r_{t}=0$, from Eq. (11)
we obtain the following equation for the
parameter $\kappa$, 
\begin{equation}
2\rho\kappa-
\varrho_{sx}k_{x}^{2}=\frac{\varrho_{sx}q^{2}\kappa}{\varrho_{sz}k_{x}^{2}}
(2\rho+\varrho_{sz}\kappa), 
\end{equation}
which has two roots in the longwavelength limit $k_{x}\ll\rho/
\sqrt{\mid\varrho_{sx}\varrho_{sz}\mid}$. In the case of 
$\varrho_{sx}\ll\varrho_{sz}$, which is
realized for the planar defect with $1-c_{l}^{(s) 2}/c_{l}^{2}\ll 1$, one of the
roots of Eq. (25) 
is close to the first root $\kappa=\kappa_{1}$ (20) of Eq. (19). Another root of Eq. (25) 
is unphysical one for such a 2D defect. Equation (25) 
can be also obtained from the solution of the problem of the total 
``nonreflection" of
the transverse acoustic wave at the planar defect for the overcritical
angle of incidence, when the potentials have the following form: 
\begin{eqnarray}
\psi^{(1)}&=&e^{iqz+ik_{x}x-i\omega t},
\quad
\varphi^{(1)}=r_{l}e^{\kappa z+ik_{x}x-i\omega t},\\
\psi^{(2)}&=&d_{t}e^{iqz+ik_{x}x-i\omega t},\quad
\varphi^{(2)}=d_{l}e^{-\kappa z+ik_{x}x-i\omega t}.
\end{eqnarray}
One can show that the wave fields (26) and (27) satisfy boundary conditions 
(5)-(7) if the parameter $\kappa$ is a solution of Eq. (25) and 
the transmission 
 amplitude $d_{t}$ in Eq. (27) has the unit modulus, $d_{t}=\exp(i\phi_{d})$. 
 From Eq. (13) we find the following expression 
for the transmission amplitude $d_{t}$ under the condition (25):
\begin{equation}
d_{t}=\frac{2\rho\kappa -\varrho_{sx}k_{x}^{2}+i\varrho_{sx}\kappa q}
{2\rho\kappa -\varrho_{sx}k_{x}^{2}-i\varrho_{sx}\kappa q}.
\end{equation}
For the solution of Eq. (25) close to the value $\kappa=\kappa_{1}$ given by
Eq. (20), from Eq. (28) follows that $d_{t}\approx 1$ (or $\phi_{d}\ll 1$). 

Therefore at a 2D defect with
$\varrho_{sx}\ll\varrho_{sz}$ (or $1-c_{l}^{(s) 2}/c_{l}^{2}\ll 1$), the resonant
conditions for the total reflection and total transmission of the transverse 
acoustic phonon through the planar defect are very close and one needs to
account for the additional (relatively weak) interactions at the 2D defect to
distinguish between these two phenomena. 
Dissipative interaction of bulk phonons with a 2D defect, caused by bulk and
surface dissipation,  is an example of 
such physically real interaction which is always present in a solid. 
(Even zero-point lattice oscillations  at
$T=0$ are related, via the fluctuation-dissipation theorem, to the 
dissipative properties of the lattice). 

Below we account for dissipation in two different ways. At first we
calculate the coefficient of surface absorption $R_{s}$  in
the limit of relatively weak absorption, using  the
dissipative function of a solid and expressions 
(12) and (14) for the reflection and 
transmissions amplitudes $r_{l}$ and $d_{l}$  
without an account for the dissipation. Then we 
calculate  the reflection and transmission
amplitudes  $r_{t}$ and $d_{t}$  with an account for dissipation and 
obtain general expression 
for the coefficient of resonant surface absorption, making use of these amplitudes.  
The coefficient of surface absorption is determined as the 
ratio between the dissipated, $E_{d}$, and incident, $E_{inc}$, (time
averaged) acoustic
energies per unit
time per unit surface area (see, e.g., Refs. [10,18,19]). 
The viscosity of the solid gives the following
contribution to the energy dissipated in the bulk of a solid, $E_{d}^{(b)}$,  and
at a surface of a 2D defect, $E_{d}^{(s)}$:
\begin{equation}
E_{d}=\int_{-\infty}^{\infty}
[\eta(\dot{u}_{ik}-\frac{1}{3}\dot{u}_{ll})^{2}+\frac{1}{2}\zeta
\dot{u}_{ll}^2]
dz+\eta^{(s)}(\dot{u}_{\alpha\beta}^{s}-\frac{1}{2}\dot{u}_{\alpha\alpha}^{s})^{2}
+\frac{1}{2}\zeta^{(s)}
\dot{u}_{\alpha\alpha}^{s 2}\equiv E_{d}^{(b)}+E_{d}^{(s)},
\end{equation}
where $\eta$ and $\zeta$ are shear and second bulk viscosities, $\eta^{(s)}$ and $\zeta^{(s)}$ are shear and second surface viscosities,
the uppex index $s$ denotes the values of the elastic displacements at the defect
plane, the Greek indices take the numbers $1$ and $2$ and numerate the
coordinate axis in the plane tangential to the 2D defect. The 
thermal conduction of the solid also gives the contribution to $E_{d}$ 
(see, e.g., Ref. [18]) which for the brevity we will not consider explicitly.
The acoustic energy, incident at the unit area of the 
planar defect per unit time by the transverse acoustic wave with amplitude
$u_{to}$ and equal to unity value of the vector potential $\psi$, 
has the following form (see Eqs. (1) and (8)):
\begin{equation}
E_{inc}=\frac{1}{2}\rho c_{t}\omega^{2}\cos\theta_{t}\mid u_{to}\mid^{2}=
\frac{1}{2}\rho c_{t}^{2}\omega qk_{t}^{2}.
\end{equation}

Now we take into account that at resonance of total reflection at a 2D defect
with $\varrho_{sx}>0$, 
the quasilongitudinal  leaky wave is excited (see Eqs. (19) and (20)). 
Since at resonance  the transmission and reflection
amplitudes  for the longitudinal wave are large, 
$\mid r_{l}\mid\approx \mid d_{l}\mid\approx k_{x}/\kappa\gg 1$, 
see Eqs. (12) and (14), this quasilocalized wave 
gives the main contribution to surface losses for 
$\theta_{t}=\theta_{t}^{(r)}$. 
Then with the use of 
Eqs. (29) and (30) together with Eqs. (2), (4), (8), (12), (14), (17), and
(20), we calculate the coefficient  of surface  absorption $R_{s}$ in
the limit of relatively weak absorption (when $R_{s}\ll 1$):
\begin{equation}
R_{s}=\frac{E_{d}}{E_{inc}}=\frac{4\rho}{\varrho_{sx}^{2}c_{l}\tan\theta_{t}^{(r)}}
\left[\frac{2\eta_{11}\rho}{\varrho_{sx}k_{x}^{2}}+\eta_{11}^{(s)}\right],
\end{equation}
where $\eta_{11}=(4/3)\eta+\zeta$ and $\eta_{11}^{(s)}
=\eta^{(s)}+\zeta^{(s)}$. 
[If one 
accounts for the contribution of thermal conduction in the solid to the 
bulk 
dissipative function $E_{d}^{(b)}$, see, e.g.,
Ref. [18], the same linear
combination of the $\eta_{11}$  and the 
coefficient of thermal conductivity, which determines the
absorption coefficient for longitudinal bulk elastic wave,  will appear 
in r.h.s. of Eq. (31) instead of
$\eta_{11}$.]  According to Eq. (31), the main contribution to surface losses 
is given by the bulk dissipation (viscosity and thermal conduction) in a
solid, 
when 
$E_{d}^{(s)}/E_{d}^{(b)}=\eta_{11}^{(s)}/(\eta_{11}\delta)\ll
1$, due to large penetration depth 
 $\delta=\kappa^{-1}=2\rho/(\varrho_{sx}k_{x}^{2})\sim \lambda^{2}/a^{\star}\gg
\lambda\gg a^{\star}$ of the quasilongitudinal leaky wave. Therefore 
the coefficient  of surface  absorption $R_{s}$ is inversely proportional
to the square of the frequency in
the limit of relatively weak absorption, and 
the account for the surface dissipation in $R_{s}$ has a sense 
only in the case of 
anomalously 
strong surface absorption (when  the 
surface dissipative length greatly exceeds the bulk one, 
$a_{l}^{(s)}\gg a_{l}$, see Eqs. (32) and (37) below). 

Now we turn to the calculation of the reflection and transmission
coefficients with an account for the dissipation which permits us to find the
coefficient of surface absorption $R_{s}$ not only in the limit of relatively weak
absorption, when $R_{s}\ll 1$, but also in the case when it reaches the value
of the order of unity. To account for the dissipation in the
reflection/refraction of elastic waves, one can use 
the ``dissipative acoustic theory" (see, e.g., Refs. [8,9,20]) in which 
the elastic moduli (and corresponding sound velocities and wavevectors) 
are assumed to be complex
quantities in bulk
equations of motion and boundary conditions for them. Since in the problem
under consideration  the main contribution to resonant 
surface absorption is
given by the quasilongitudinal (leaky) wave, 
below we consider the longitudinal, 
bulk and surface, 
sound velocities as
\begin{equation}
c_{l}=c_{lo}-i\omega a_{l}, ~~
c_{l}^{(s)}=c_{lo}^{(s)}-i\omega a_{l}^{(s)}.
\end{equation}
In Eq. (32) and in the following, 
index $o$ refers to the limit in which one neglects the dissipation. 
The parameter  $a_{l}$ in Eq. (32)  is a bulk longitudinal ``dissipative
length" which is directly related to the absorption coefficient 
$\gamma_{l}$ of the longitudinal bulk acoustic wave: 
$a_{l}=\gamma  _{l}c_{lo}^2/\omega^2$ (and 
$a_{l}=\eta_{11}/2\rho c_{lo}$ if one does
not account explicitly for the thermal conduction in the solid, cf. Eqs. (29),
(31) and
Ref. [18]). The damping of the transverse elastic waves can be  also taken into
account by introducing, similarly to
Eq. (32), the  transverse dissipative length $a_{t}$.   
For the low frequencies $\omega\tau\ll 1$ (where $\tau$ is an 
effective relaxation time of the excitations which cause the elastic wave
absorption),  one has $\gamma_{l,t}\propto\omega^{2}$  and the
lengths $a_{l,t}$ do not depend on the frequency and are  proportional to the effective mean free
path $l$  of the excitations. For the high frequencies $\omega\tau\gg 1$, in
single crystals 
one has $\gamma_{l,t}\propto\omega$  and the
lengths $a_{l,t}$ are inversely proportional to the frequency:  
$a_{l,t}=p_{l,t}c_{lo,to}/\omega$, where the dimensionless parameters $p_{l,t}$ 
do not depend on the frequency and the effective mean free path $l$ 
and describe phenomenologically the absorption of bulk phonons in the ballistic
regime (cf. Refs. [8,9,11]). 
In both regimes, in the hydrodynamic $\omega\tau\ll 1$
and ballistic $\omega\tau\gg 1$ ones, 
relatively weak sound absorption in a
solid corresponds to $\omega a_{l,t}\ll c_{lo,to}$ (or $p_{l,t}\ll 1$). 
The dissipative lengths $a_{l,t}$ can vary in rather broad
interval of values 
and are functions of the state of the solid (its temperature, impurity
concentration, etc.). These lengths strongly depend on the type of the solid
(metal, insulator, or semiconductor), and usually $a_{l}>a_{t}$.  
In crystal quartz these lengths are rather short, for instance  
$a_{l}\sim 6 \AA$ both at room
temperature [21] and at $T=140K$ [22]. In  metals whose Fermi surface can be
approximated by the free-electron spherical Fermi surface (such as copper, sodium, lead, tin, and indium), the electron 
viscosity $\eta_{e}$ is proportional to the electrical conductivity and one
has $\eta_{e}\sim p_{F}N_{e}l_{e}$, where $p_{F}$, $l_{e}$  and $N_{e}$ are 
the Fermi boundary momentum, electron mean free path
 and the number of conduction 
electrons per unit volume, respectively. In this case for the low frequencies one has  
$a_{l}\sim (p_{F}/M_{i}c_{l})l_{e}$, 
where $M_{i}$ is a mass of an ion in the metal. In particular, for the copper 
one has $a_{l}\sim 2\cdot 10^{-3} l_{e}$, and 
therefore for the low frequencies 
the longitudinal dissipative length $a_{l}$ can have the order of
a micron in pure enough copper in the limit
of $T=0$ (when $l_{e}$ reaches $10^{-2}$ $cm$, see, e.g.,
Ref. [23]). 

The parameter $a_{l}^{(s)}$ in Eq. (32) is the surface (longitudinal) 
dissipative length which is  
related to the surface  
viscosity $\eta_{11}^{(s)}=\eta^{(s)}+\zeta^{(s)}$ (see Eqs. (29) and (31)): 
$a_{l}^{(s)}=\eta_{11}^{(s)}/(2\varrho_{s}c_{lo}^{(s)})$. The surface dissipative
length enters the considered problem via the complex elastic modulus 
$\tilde{h}_{11}=\varrho_{s}c_{l}^{(s)2}$ in the boundary condition
(7). Interface roughness and near-surface lattice imperfections 
increase the 
damping of acoustic phonons in the vicinity of 2D defect. 
It means that effectively one can have    
$a_{l}^{(s)}\gg a_{l}$, which justifies  in this case  
the account for the surface (interface) dissipation in 
addition to the bulk one. Such strong inequality between surface (interface)
and bulk dissipative lengths 
can occur also 
for a  2D electron gas at an interface between two  similar semiconductors (such as 
$GaAs/AlGaAs$ heterojunction), since in (semi)conductors at low enough temperature these 
lengths for the low frequencies are proportional to the local electrical conductivity (see also Ref. [5]).     

Even with an account for the dissipation in the 
reflection/refraction of elastic waves, we can still consider the 
$k_{x}$ component of the incident wavevector
as a real quantity. Then making use of Eq. (32) and assuming that 
$a_{l}^{(s)}\gg a_{l}$, we obtain the following expansions: 
\begin{eqnarray}
\kappa&=&\sqrt{k_{x}^{2}-\frac{\omega^{2}}{c_{l}^{2}}}
=\kappa_{o}-ia_{l}\frac{\omega^{3}}{c_{lo}^{3}\kappa_{o}}, \\
\varrho_{sx}&=&\varrho_{s}\left[1-\frac{c_{l}^{(s) 2}}{c_{l}^{2}}\right]
=\varrho_{sxo}+2ia_{l}^{(s)}\frac{\omega
\varrho_{s}c_{lo}^{(s)}}{c_{lo}^{2}},
\end{eqnarray}
which are valid for $(a_{l},a_{l}^{(s)})k_{x}\ll 1$ and 
$a_{l}k_{x}^{3}\ll\kappa_{o}^{2}$ (or
$a_{l}\ll\varrho_{sxo}^{2}k_{x}/\rho^{2}\sim a^{\star 2}k_{x}$).  
[At resonance with the quasilongitudinal leaky wave, the account for the
imaginary part of the $q$ component of the incident wavevector  gives only
relatively weak corrections, of order $a_{t}k_{x}\ll 1$, to the final
expressions,  see Eqs. (35)-(37) below]. 
Making use of Eqs. (33), (34), from Eqs. (11) and (13) we find 
the reflection and 
transmission coefficients  $r_{t}^{(r)}$ and $d_{t}^{(r)}$ for the
transverse bulk phonon at resonance with the leaky wave 
at a 2D defect with an acount for the dissipation:
\begin{eqnarray}
r_{t}^{(r)}&=&\frac{\varrho_{sxo}^{3}c_{lo}\tan\theta_{t}^{(r)}k_{x}^{2}}
{4\rho(2a_{l}\rho^{2}c_{lo}
+a_{l}^{(s)}\varrho_{sxo}c_{lo}^{(s)}\varrho_{s}k_{x}^{2}) 
+\varrho_{sxo}^{3}c_{lo}\tan\theta_{t}^{(r)}k_{x}^{2}}, \\
d_{t}^{(r)}&=&\frac{4\rho(2a_{l}\rho^{2}c_{lo}
+a_{l}^{(s)}\varrho_{sxo}c_{lo}^{(s)}\varrho_{s}k_{x}^{2})}
{4\rho(2a_{l}\rho^{2}c_{lo}
+a_{l}^{(s)}\varrho_{sxo}c_{lo}^{(s)}\varrho_{s}k_{x}^{2}) 
+\varrho_{sxo}^{3}c_{lo}\tan\theta_{t}^{(r)}k_{x}^{2}}.
\end{eqnarray}
[Using expansions (33), (34) and the corresponding expansion for the $q$
wavevector component, by equating to zero each of the two brackets in 
the denominator $\Delta$ (15) of all reflection/refraction
amplitudes one can also find with an account for dissipation the dispersion
relations for two ``sagittal" waves, with quasilongitudinal and quasitransverse 
polarizations, which can propagate along
the 2D defect, cf. Refs. [1,2]]. 
With the help of Eqs. (35) and (36) one can readily obtain the coefficient of 
surface absorption $R_{s}$ as a dimensionless difference between the incident 
and reflected/transmitted fluxs of acoustic energy:
\begin{equation}
R_{s}=1-\mid r_{t}\mid^2-\mid d_{t}\mid^2=
\frac{8(2a_{l}\rho^{2}c_{lo}
+a_{l}^{(s)}\varrho_{sxo}c_{lo}^{(s)}\varrho_{s}k_{x}^{2})\rho
\varrho_{sxo}^{3}c_{lo}\tan\theta_{t}^{(r)}k_{x}^{2}}
{[4\rho(2a_{l}\rho^{2}c_{lo}
+a_{l}^{(s)}\varrho_{sxo}c_{lo}^{(s)}\varrho_{s}k_{x}^{2}) 
+\varrho_{sxo}^{3}c_{lo}\tan\theta_{t}^{(r)}k_{x}^{2}]^{2}}.
\end{equation}

Expressions (35)-(37)  represent the main result of this work. The results of
Refs. [6,7], namely $r_{t}^{(r)}=1$,  $d_{t}^{(r)}=0$, follow from Eqs. (35),
(36) in the limit in which one neglects the dissipation: $a_{l}=a_{l}^{(s)}=0$. 
If the introduced   
relations  between the 
dissipative lengths and viscosities are used
in Eq. (37), namely $2a_{l}\rho c_{lo}=\eta_{11}$ and $2a_{l}^{(s)}\varrho_{s}
c_{lo}^{(s)}=\eta_{11}^{(s)}$, the expression (37) for $R_{s}$ coincides 
exactly with the expression (31) in the limit of relatively weak absorption. 
From Eqs. (31), (35)-(37) follows 
that this limit corresponds to extremely weak dissipation: 
$a_{l}\ll\varrho_{sx}^{3}\tan\theta_{t}^{(r)}k_{x}^{2}/8\rho^{3}\sim a^{\star
3}k_{x}^{2}\ll a^{\star}$ and
$a_{l}^{(s)}\ll\varrho_{sx}^{2}/\varrho_{s}\rho$. 
Only in this limit one has $\mid d_{t}^{(r)}\mid\ll
1$, $r_{t}^{(r)}\approx 1$, and $R_{s}\ll 1$. But in view of the 
abovementioned characteristic values of the bulk dissipative length $a_{l}$,
especially in metals in which $a_{l}$ can have the order of a micron, 
this limit 
is hard to reach  experimentally because of the assumed smallness of the
dimensionless parameter $a^{\star}k_{x}$: one
needs $\sqrt{a_{l}/a^{\star}}\ll a^{\star}k_{x}\ll 1$. 
[Moreover, the effective thickness of the layer 
$a^{\star}=\varrho_{sx}/\rho=h(1-c_{l}^{(s) 2}/c_{l}^{2})$,  
which enters Eqs. (35)-(37),  is much smaller than the ``actual" thickness 
$h=\varrho_{s}/\rho$ of the layer if $1-c_{l}^{(s) 2}/c_{l}^{2}\ll 1$]. 
Therefore from the experimental point of view most plausible is the opposite limit 
$a_{l}\gg\varrho_{sx}^{3}\tan\theta_{t}^{(r)}k_{x}^{2}/8\rho^{3}
\sim a^{\star
3}k_{x}^{2}$ (which is
consistent with the initial assumptions of the proposed approach that 
$a_{l}\ll\varrho_{sx}^{2}k_{x}/\rho^{2}\sim a^{\star 2}k_{x}$ and 
$a^{\star}k_{x}\ll 1$, see Eq. (33)). In this low-frequency limit 
bulk phonons traverse  the 2D defect almost without reflection and
surface absorption, when $r_{t}\ll 1$, $d_{t}\approx 1$, $R_{s}\ll 1$  
for all angles of incidence, including the resonant
one (as it should be from the ``intuitive" point of view).  
With the use of Eqs. (11), (13), (33), and (34) one can also 
ascertain that the dissipation does not considerably change 
the reflection and transmission coefficients for the transverse elastic wave at
resonance of $total$ $transmission$, described by Eq. (25), in contrast to the
reflection/transmission coefficients at resonance of $total$ $reflection$,
described by Eq. (19). The total reflection of the grazing
acoustic wave at a 2D defect layer [1,2] and the total reflection of a bulk
phonon at resonance with an 
asymmetric vibrational mode of an  ``inhomogeneous" 2D elastic layer with complex
internal structure [5] are also not affected strongly by
the bulk dissipation since these phenomena are 
not accompanied by the resonant excitation of the 
deeply penetrating leaky wave. 
It means that resonant  interaction with the deeply penetrating leaky wave 
and total reflection of the 
long transverse elastic wave at a ``homogeneous" planar defect, 
described in 
Refs. [6,7] without an account for the dissipation, turns to be extremely 
sensitive to the dissipative losses in a solid, especially to the bulk
ones.  From the experimental point of view, an interest can   
represent an ``intermediate" case when   
 $a_{l}\sim\varrho_{sx}^{3}\tan\theta_{t}^{(r)}k_{x}^{2}/8\rho^{3}\sim a^{\star
3}k_{x}^{2}$. Indeed, from Eq. (37) follows that if  
$a_{l}=(\varrho_{sx}^{3}c_{l}\tan\theta_{t}^{(r)}-
4\varrho_{s}\varrho_{sx}\rho 
a_{l}^{(s)}c_{l}^{(s)})k_{x}^{2}/8\rho^{3}c_{l}>0$,
the coefficient of surface absorption $R_{s}$ reaches it maximal value $0.5$
(when $r_{t}^{(r)}=d_{t}^{(r)}=0.5$). It means that one half of
the energy flux of the incident long acoustic wave  is absorbed by a thin, 
with $a^{\star}\ll\lambda$, elastic layer (2D defect)  sandwiched  in a
solid. This effect of anomalous surface absorption is similar to 
anomalous
absorption, also with $R_{s}=0.5$, of the grazing shear elastic wave 
in a thin gap between two solids caused by the dissipative van der 
Waals interaction between the solids [13]. In both cases,  the anomalous
surface (interface) absorption of the bulk phonon is stipulated by resonant
dissipative 
interaction with the leaky elastic wave. 

In conclusion, we predict an extreme sensitivity 
to the dissipative losses, especially to the bulk
ones, of  the resonant  
interaction of  bulk phonons with a thin elastic layer (2D defect) sandwiched
in a solid. The coefficients of the resonant transmission and reflection of
the 
transverse phonon at the 2D defect are
derived with an account for the bulk and surface dissipation. We show that 
almost total reflection of the transverse phonon at a 2D defect, described in
Refs. [6,7] without an account for the dissipation,  
occurs only in the limit of extremely  weak dissipation  when 
$a_{l}\ll a^{\star 3}k_{x}^{2}$ and
$a_{l}^{(s)}\ll\varrho_{sx}^{2}/\varrho_{s}\rho$, which is hard to realize 
experimentally.  In this limit the coefficient of surface absorption $R_{s}$  of the incident 
phonon is small and is 
proportional to the bulk viscosity and thermal conductivity of the solid, 
but is inversely proportional to the
square of the frequency. We show that almost 
total resonant reflection of the transverse phonon at the 2D defect can be 
 changed into
almost total transmission by relatively weak bulk 
absorption. Anomalous surface absorption of the transverse phonon, 
when $R_{s}$ reaches its maximal
value $0.5$ and which is caused by resonant
dissipative 
interaction with the leaky elastic wave, is
predicted for the case of intermediate bulk dissipation when 
$a_{l}\sim a^{\star 3}k_{x}^{2}$.
     
The authors are grateful to A.N. Darinskii, R. Heid, A.M. Kosevich, A. Mayer, 
V.G. Mozhaev  and K.N. Zinov'eva 
for the useful discussions at different stages of the 
work. Yu.A.K. acknowledges the support from the Max Planck
Society. E.S.S. gratefully acknowledges the
Max Planck Institute for Physics of Complex Systems in Dresden 
for hospitality during the preparation of this work.

\end{document}